\documentclass[10pt,english]{iopart}
\pdfoutput=1
\usepackage{iopams}
\usepackage{setstack}
\usepackage{amstext}
\usepackage{babel}
\usepackage{graphicx}
\usepackage{bm,amssymb}
\usepackage{color}
\usepackage{esint}

\begin{document}

\title{Strong light-matter interaction in systems described by a modified Dirac equation}
\author{N. M. R. Peres and Jaime E. Santos}

\address{Physics Department and CFUM, University of Minho, P-4710-057, Braga, Portugal}
\ead{peres@fisica.uminho.pt}
\date{\today}

\begin{abstract}The bulk states of some materials, such as topological insulators, are described by a
modified Dirac equation. Such systems may have trivial and non-trivial phases. In this paper,
we show that in the non-trivial phase a strong light-matter interaction exists in a two-dimensional
system, which leads to an optical conductivity at least one order of magnitude larger than that of
graphene.
\end{abstract}

 
\maketitle

\section{Introduction}

Light-matter interaction is a central research topic in atomic, particle, and
condensed matter physics. 
In the solid state context \cite{OpticalBook}, optical spectroscopy of materials is a powerful 
method of gaining  information on the dynamics of electrons in a given material.
In general, one is interested in the optical properties of 
many different types of systems: superconductors, ordinary metals,
semiconductors, insulators, two dimensional systems, 
such as graphene \cite{rmp,rmpPeres} and dichalcogenides,
topological insulators \cite{SCZhang,Manoharan,PWKane,TopBook}, and others.

Topological insulators are characterized by being insulators in the bulk (at least ideally)
and conducting at the surface. The effective low energy Hamiltonian describing the 
helical (in two dimensions) and surface (in three dimensions) states is the massless Dirac
equation \cite{SCZhangNatPhys,rmpKane,rmpSCZ}. On the other hand, the low energy Hamiltonian
of the bulk states of a topological insulator can be approximated by a modified Dirac equation,
with a mass term that is momentum dependent. 

The optical conductivity of the surface states of topological insulators has been recently
considered \cite{carbote}. It was found that due to important hexagonal warping the optical
conductivity of that class of topological insulators (Bi$_{2}$Te$_{3}$) deviates considerably
from the value predicted and measured for neutral graphene \cite{nair,StauberGeim}:
\begin{equation}
 \sigma_{xx}(\omega)=\frac{\pi e^2}{2h}\equiv\sigma_0\,.
\end{equation}
The effect of disorder on the optical conductivity of the surface states of the 
topological insulator  Bi$_2$Se$_3$ has recently been studied \cite{ZieglerTI}.
The optical conductivity of  Bismuth-base topological insulators has been
experimentally investigated \cite{BismuthBasov,Bismuth,OpticalBi2Te2Se}.
 
On the other hand, and to the best of our knowledge, the optical 
conductivity of the bulk states of topological insulators
has not been studied theoretically. This is understandable, 
since the focus has been on the dissipationless nature
of the edge states, which can propagate without scatter off impurities.
A topological insulator may or may note
have edge states depending on the value of the Chern number. 
If this quantity is finite then there will be edge states
and the system is said to be non-trivial. On the other hand, 
if the Chern number is zero there will be no edge states
and the system is said to be trivial.
In what follows, we will thus consider the contribution of the bulk states to the 
optical conductivity of a topological insulator, described by a modified Dirac 
equation in two dimensions. 
We will see that there is regime of parameters, where the Chern number 
is finite, which show a clear signature of the non-trivial nature of the system, 
in the sense defined above.

\section{The modified Dirac equation and the density of states}
The most general two-band model in two dimensions has the form
\begin{equation}
 H=d_x(\bm k)\sigma_x+d_y(\bm k)\sigma_y+d_z(\bm k)\sigma_z\,,
\label{eq_H0}
\end{equation}
where $\sigma_i$ is the $i=x,y,z$ Pauli's matrix.
We shall consider a particular case where  $d_x(\bm k)=v\hbar k_x$, $d_y(\bm k)=v\hbar k_y$, and 
$d_z(\bm k)=M(\bm k)$ --
that is the case of a Dirac Hamiltonian; we also assume that the mass
term is momentum dependent. 
Such model is termed the modified Dirac equation \cite{SPINShen} and applies to 
the bulk states of 
topological insulators around the $\Gamma-$point of the Brillouin zone.
The eigenvalues of this Hamiltonian are given by
\begin{equation}
E_{\bm k,\lambda}=\lambda\sqrt{v^2\hbar^2k^2+M^2(\bm k)}\,, 
\label{eq_eigenv}
\end{equation}
 with $k=\sqrt{k_x^2+k_y^2}$
and $\lambda=\pm1$. 
The normalized eigenstates are
\begin{equation}
 \psi_{\bm k,+}=\frac{1}{\sqrt{2E_{\bm k,+}}}\left(
\begin{array}{c}
 \sqrt{E_{\bm k,+}+M(\bm k)}e^{-i\theta}\\
\sqrt{E_{\bm k,+}-M(\bm k)}
\end{array}
\right)
\end{equation}
and
\begin{equation}
 \psi_{\bm k,-}=
\frac{1}{\sqrt{2E_{\bm k,+}}}
\left(
\begin{array}{c}
 -\sqrt{E_{\bm k,+}-M(\bm k)}e^{-i\theta}\\
\sqrt{E_{\bm k,+}+M(\bm k)}
\end{array}
\right)\,,
\end{equation}
where $\theta=\arctan(k_y/k_x)$. We further consider that the mass term has the form
$M(\bm k)=mv^2-B\hbar^2k^2$, with $B$ and $m$ constants that can be either positive or negative.
Such type of model appears in the theory of topological insulators \cite{rmpSCZ}.

If we define the vector $\bm d=[d_x(\bm k),d_y(\bm k),d_z(\bm k)]$, the Chern number has the 
form \cite{SCZhang}
\begin{eqnarray}
 n_c=\frac{1}{4\pi}\int_{\mbox{BZ}}dp_xdp_y
\frac{1}{E^3_{\bm k,+}}\left(
\frac{\partial \bm d}{\partial p_x}\times\frac{\partial \bm d}{\partial p_x}
\right)\cdot \bm d\,,
\end{eqnarray}
where $p_j=\hbar k_j$ and the integral runs over the full Brillouin zone.
For our model Hamiltonian,
the Chern number acquires the form
\begin{eqnarray}
 n_c&=& \frac{1}{2}v^2\int_0^\infty \frac{p(Bp^2+mv^2)dp}{[v^2p^2 + (mv^2-Bp^2)^2]^{3/2}}
\nonumber\\
&=&
\frac{1}{2}[\mbox{sgn\,}(m)+\mbox{sgn\,}(B)]\,.
\end{eqnarray}
We then conclude that when $mB>0$ there is a Hall current and the system is said to be 
topologically non-trivial; when $mB<0$, $ n_c=0$ and the system is trivial.

\begin{figure}[ht]
\includegraphics*[width=8cm]{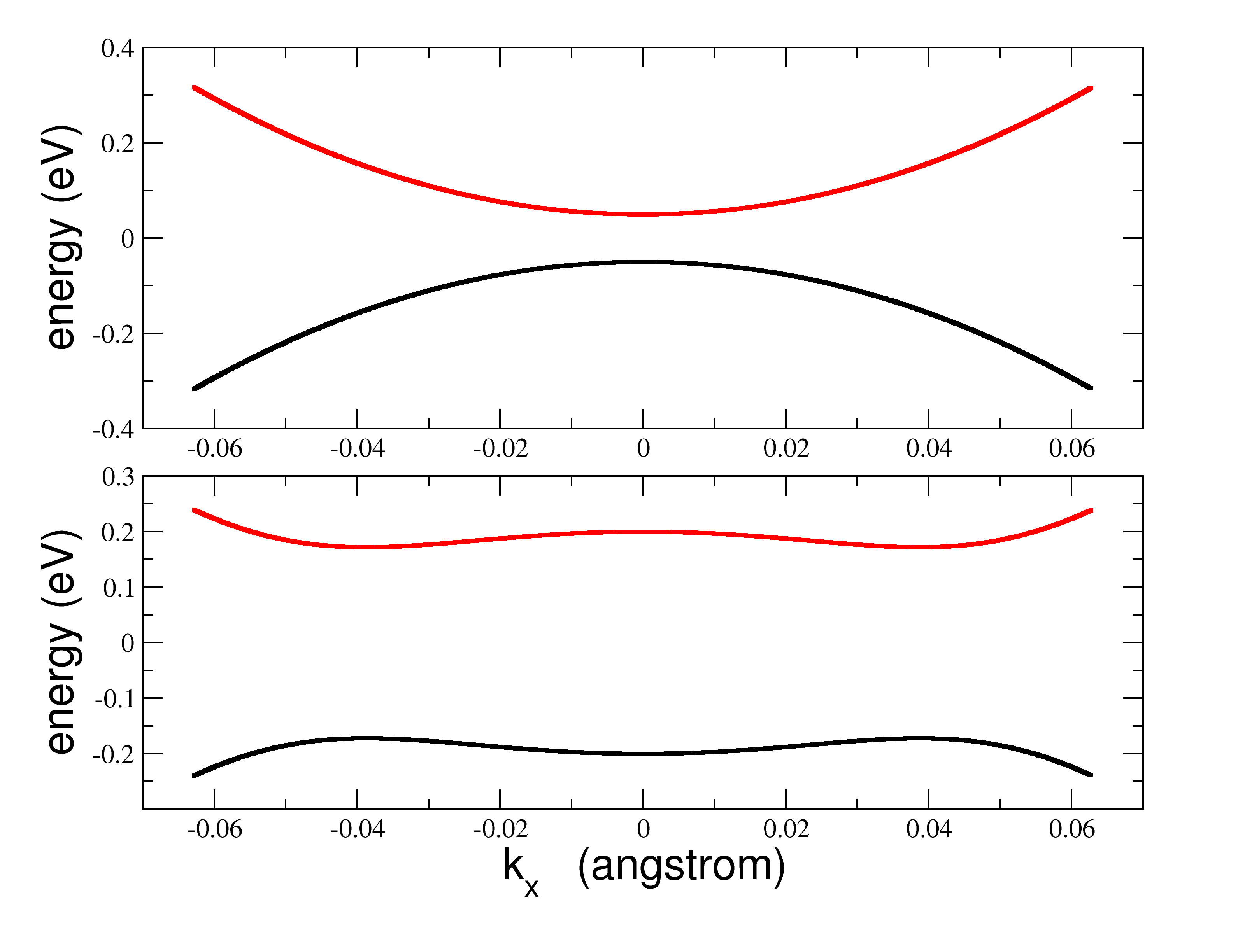}
\caption {Band structure  of the modified Dirac equation, as defined by 
Eq. (\ref{eq_eigenv}), for $k_y=0$ as function of $k_x$. 
The used parameters are
$v\hbar=3.65$ eV$\cdot$\AA, $B\hbar^2$=-68 eV$\cdot$\AA$^2$, as for 
HgTe quantum wells. For  $mv^2$ we take the value -0.05 eV for the upper panel
and -0.2 for the lower panel. In the latter case the system is in the regime 
$2Bm>1$ and a Mexican-hat type of band is seen.}
\label{Fig_BS}
\end{figure}

In Fig. \ref{Fig_BS}, we depict the band structure of the modified Dirac equation in the 
regimes $2Bm\leq 1$ and $2Bm>1$ separately. It is clear that in the latter case 
the gap is off the $\Gamma-$point. 
The band structure has, in this case, a Mexican hat shape. Then, there is a full 
circumference in momentum space where the group velocity is zero and this has
consequences in the density of state, as we shall see below. Although we are using
for the parameters $v\hbar$ and $B\hbar^2$ those of HgTe quantum wells we make no claim that
our results are directly applicable to that particular system, since the value we use for
$mv^2$ is different from what is reported in the literature \cite{rmpSCZ}. We simply fix 
$mv^2$ to a value that places the system in the regime $2Bm>1$.

The different behaviour of the system when $Bm$ is either positive or negative can also 
be seen from the average value of the spin operator, defined as
\begin{equation}
 \bm s = (\sigma_x,\sigma_y,\sigma_z)\,.
\end{equation}
The several terms are 
\begin{eqnarray}
 \langle \psi_{\bm k,\pm} \vert \sigma_x\vert \psi_{\bm k,\pm}\rangle 
&=& \pm \frac{k_xv\hbar}{E_{\bm k+}}\,,
\nonumber\\
 \langle \psi_{\bm k,\pm} \vert \sigma_y\vert \psi_{\bm k,\pm}\rangle 
&=& \pm \frac{k_yv\hbar}{E_{\bm k+}}\,,
\nonumber\\
 \langle \psi_{\bm k,\pm} \vert \sigma_z\vert \psi_{\bm k,\pm}\rangle 
&=& \pm \frac{mv^2-B\hbar^2k^2}{E_{\bm k+}}\,.
\end{eqnarray}
\begin{figure}[ht]
\includegraphics*[width=8cm]{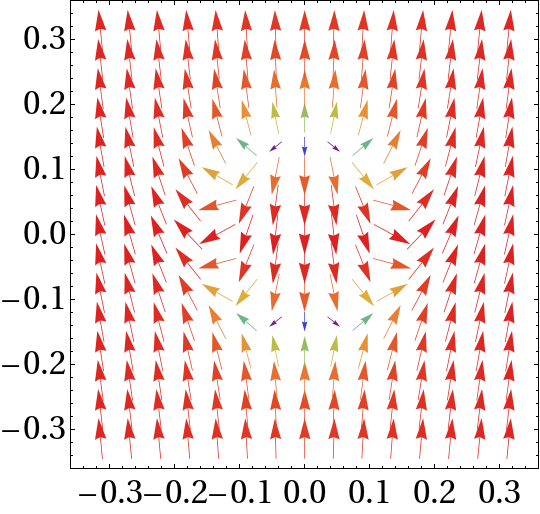}
\caption {
Vector plot of
$\langle \psi_{\bm k,+} \vert \sigma_x\vert \psi_{\bm k,+}\rangle$
and
 $\langle \psi_{\bm k,+} \vert \sigma_z\vert \psi_{\bm k,+}\rangle$
as function of $k_x$ and $k_y$, expressed in \AA.
 The used parameters are
$v\hbar=3.65$ eV$\cdot$\AA, $B\hbar^2$=-68 eV$\cdot$\AA$^2$, as for 
HgTe quantum wells. For  $mv^2$ we take the value -1.5 eV, which puts the system
in the regime $Bm>0$.}
\label{Fig_Arrows_Bm_positive}
\end{figure}
In Fig. \ref{Fig_Arrows_Bm_positive} we depict a vector plot of 
$\langle \psi_{\bm k,+} \vert \sigma_x\vert \psi_{\bm k,+}\rangle$
and
 $\langle \psi_{\bm k,+} \vert \sigma_z\vert \psi_{\bm k,+}\rangle$
as function of $k_x$ and $k_y$, in the regime $Bm>0$. At the center of the Brillouin zone
the spin points down whereas when we move off the center the spin rotates and 
at large momentum it points up; in the valence band the orientation of the spin 
is the opposite.

The situation is different when we consider the regime $Bm<0$, as shown in 
\begin{figure}[ht]
\includegraphics*[width=8cm]{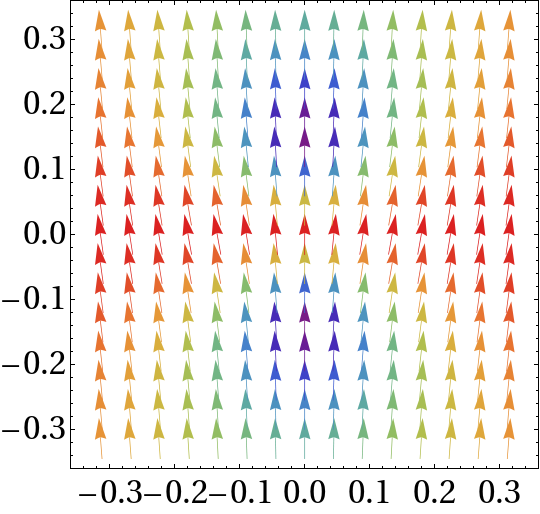}
\caption {
Vector plot of
$\langle \psi_{\bm k,+} \vert \sigma_x\vert \psi_{\bm k,+}\rangle$
and
 $\langle \psi_{\bm k,+} \vert \sigma_z\vert \psi_{\bm k,+}\rangle$
as function of $k_x$ and $k_y$, expressed in \AA.
 The used parameters are
$v\hbar=3.65$ eV$\cdot$\AA, $B\hbar^2$=-68 eV$\cdot$\AA$^2$, as for 
HgTe quantum wells. For  $mv^2$ we take the value 1.5 eV, which puts the system
in the regime $Bm<0$.}
\label{Fig_Arrows_Bm_negative}
\end{figure}
Fig. \ref{Fig_Arrows_Bm_negative}. In this case, the spin points along the same direction
independent of the position in the Brillouin zone. The different behaviour of the 
spin, depending on the sign of the product $Bm$, is a manifestation of the trivial or 
non-trivial nature of the system.

The density of states (DOS) of the conduction band of the topological insulator defined by 
Hamiltonian \ref{eq_H0} can be easily computed. This quantity is defined by the expression
\begin{equation}
\rho(\omega)=\frac{1}{A}\sum_{\bm k}\,\delta(\hbar\omega-E_{\bm k,+})\,,
\label{eq_DOSdef}
\end{equation}
where $A$ is the area of the system and $E_{\bm k,+}$ is given by \ref{eq_eigenv}
 (with $\lambda=1$). Note that the expression for the DOS of the valence 
band can simply be obtained from the one above by replacing $\omega$ with
$-\omega$, as $E_{\bm k,-}=-E_{\bm k,+}$. Converting the summation above in an
integral over the Brillouin zone in the thermodynamic limit, and performing the 
angular integral, one obtains
\begin{equation}
\rho(\omega)=\frac{1}{4\pi}\,\int_0^\infty\,du\,\delta(\hbar\omega-E_{u,+})\,,
\label{eq_DOS_2}
\end{equation}
where we have performed the substitution $u=k^2$ in the integral over the modulus
of the wave-vector, and where $E_{u,+}=\sqrt{v^2\hbar^2u+(mv^2-B\hbar^2u)^2}$.

The computation of the explicit expression for the DOS from \ref{eq_DOS_2}
can be performed by determining the values of $u$ for which the argument of 
the delta function is zero in the expression above. Thus, we need to 
determine the roots of the equation $\hbar\omega=E_{u,+}$.
Such roots will only contribute to the integral in \ref{eq_DOS_2} 
if they are real and positive. 

We start by noting that from Eq. (\ref{eq_eigenv}), we find that 
the minimum of the conduction band takes place at a momentum $k_\Delta$ 
given by
\begin{equation}
k_\Delta = v\frac{\sqrt{-1+2Bm}}{\sqrt 2 \vert B\vert  \hbar}\,,
\end{equation}
which implies that only for $2Bm>1$ does the minimum occurs off the 
$\Gamma-$point of the Brillouin zone.
In this case, the band gap is  
\begin{equation}
\Delta=\frac{v^2}{\vert B\vert}\sqrt{4Bm-1}\,,
\end{equation}
and the dispersion resembles a Mexican hat, as is clearly seen in Fig. \ref{Fig_BS}.  
If $2Bm<1$, the band gap occurs at the $\Gamma-$point and is given by 
$\Delta_\Gamma=2mv^2$. If the condition $2Bm>1$ is met, we always 
have $\Delta<\Delta_\Gamma$.

\begin{figure}[ht]
\includegraphics*[width=8cm]{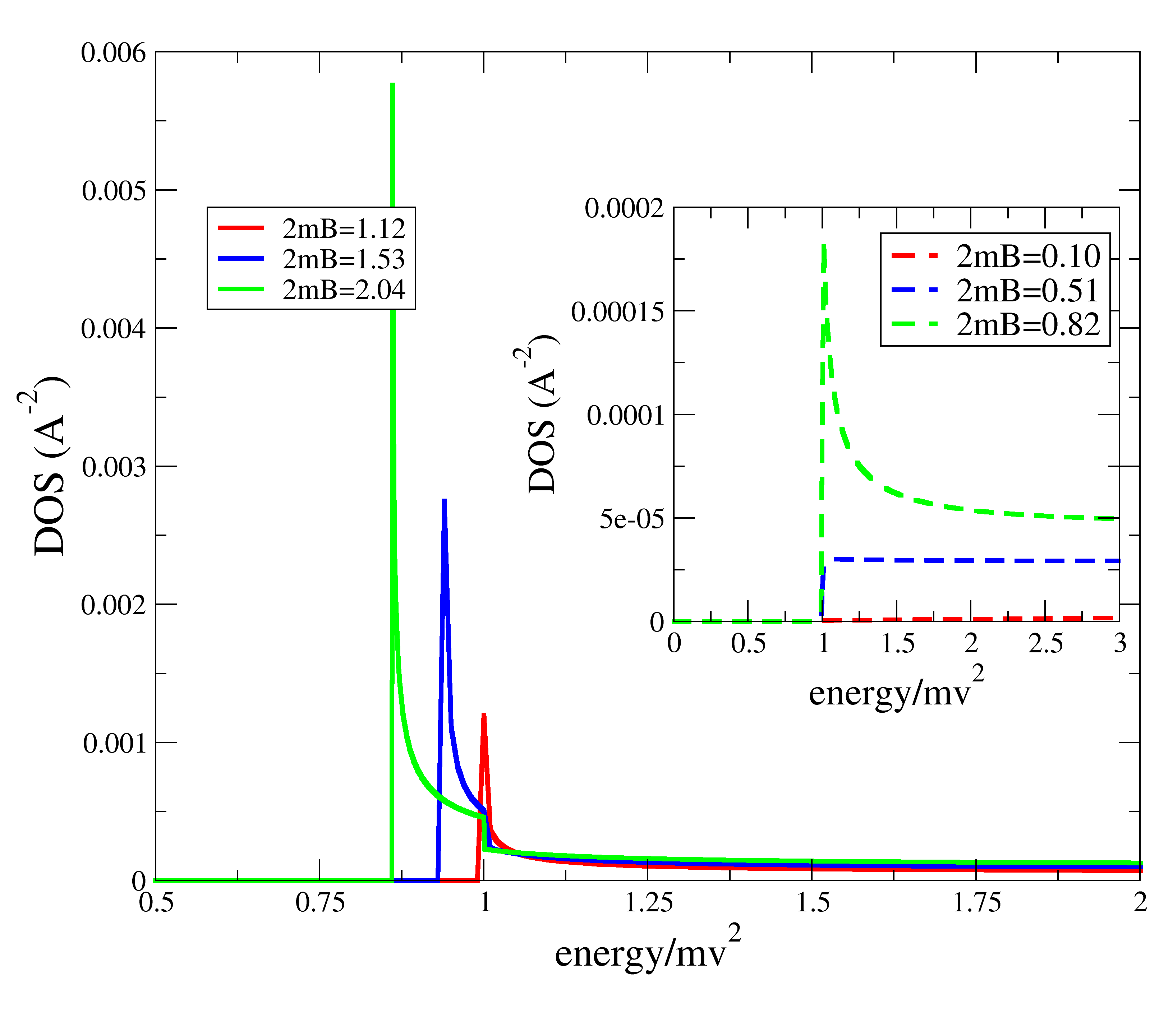}
\caption {DOS of a modified Dirac equation, as given by equation \ref{eq_DOS}. The function was scaled
by a factor of $mv^2$ so that the area under the curve is preserved with respect to the result given
in \ref{eq_DOS}, given that the energy is expressed here in units of $mv^2$ as well. The used parameters are
$v\hbar=3.65$ eV$\cdot$\AA, $B\hbar^2$=-68 eV$\cdot$\AA$^2$, as for HgTe quantum wells.}
\label{Fig_DOS}
\end{figure}

Taking into account the two different regimes $Bm<0$ and $Bm>0$ discussed above,
and the appearance of a minimum of off the $\Gamma-$point for $2Bm>1$, we need 
to consider three different cases when analysing the roots of the equation 
$\hbar\omega=E_{u,+}$:
(i) the trivial case, where $Bm<0$; (ii) the non-trivial case, where $Bm>0$ 
and $2Bm<1$; (iii) the non-trivial case, where $2Bm>1$. Defining $f(u)=\omega\hbar
-E_{u,+}$, the zeros of $f(u)$, that is $f(u_0)=0$, are 
\begin{equation}
u_0=\left\{
\begin{array}{c}
u_-=-\frac{v^2(1-2Bm)}{2B^2\hbar^2}-\frac{g(B,m,\omega)}{2B^2\hbar^2}\\
u_+=-\frac{v^2(1-2Bm)}{2B^2\hbar^2}+\frac{g(B,m,\omega)}{2B^2\hbar^2}\\
\end{array}
\right.\,,
\end{equation}
where $g(B,m,\omega)$ is defined by 
\begin{equation}
 g(B,m,\omega)=\sqrt{v^4(1-4Bm)+4B^2\omega^2\hbar^2}\,.
\label{eq_g}
\end{equation}
Since the discriminant of the square root [in $g(B,m,\omega)$] has to be positive, we find that $\hbar\omega$ can have any positive
value in both the trivial case and  the non-trivial case if $4Bm<1$. For the non-trivial case when
$4Bm>1$ we find that 
\begin{equation}
 \hbar\omega \ge \frac{v^2}{2\vert B\vert}\sqrt{4Bm-1}=\frac{\Delta}{2}\,.
\end{equation}
If $u_-$ and $u_+$ are to contribute to the integration of the $\delta-$function,
both have to be positive numbers.
This imposes some restrictions  on the values of $\hbar\omega$
depending on the  $2Bm$ parameter. A detailed analysis reveals the 
following conclusions. In the trivial case, only $u_+$ is positive and 
therefore $u_-$ does not contributes to 
the integral. In this case, we find that 
$\hbar\omega$ has to satisfy the condition
$\hbar\omega>mv^2=\frac{\Delta_\Gamma}{2}$. 
In the non trivial case, we have two regimes to consider: when (i) $2Bm<$1 and when (ii) $2Bm>1$.
In case (i), only the root $u_+$ contributes and the frequency $\omega$ has to satisfy the 
condition $\hbar\omega>\frac{\Delta_\Gamma}{2}$. In case (ii), both roots contribute. The root
$u_-$ gives a contribution in the
energy range $\frac{\Delta}{2}<\hbar\omega<\frac{\Delta_\Gamma}{2}$, whereas the root $u_+$ gives a contribution in the region $\hbar\omega>\frac{\Delta}{2}$. Taking such information into account when
computing the integral in \ref{eq_DOS_2}, we obtain, using the properties of the delta function,
the result
\begin{equation}
\rho(\omega)=\left\{
\begin{array}{c}
\frac{\omega\,\theta(\omega-\Delta_\Gamma/2)}{4\pi\hbar\sqrt{v^4(1-4mB)+4B^2\omega^2\hbar^2}}
\;\;\mbox{if}\;\;2mB<1\\
\frac{\omega\,(1+\tilde{\theta}(\Delta_\Gamma/2-\omega))\,\theta(\omega-\Delta/2)}{4\pi\hbar\sqrt{v^4(1-4mB)+4B^2\omega^2\hbar^2}}\,
\;\;\mbox{if}\;\;2mB\geq 1
\\
\end{array}
\right.\,,
\label{eq_DOS}
\end{equation}
where $\theta(x)=0$, if $x<0$, $\theta(x)=1/2$, if $x=0$, and $\theta(x)=1$, if $x>0$,
and where $\tilde{\theta}(x)=0$, if $x<0$, and $\tilde{\theta}(x)=1$, if $x\geq0$. A plot of this
quantity for selected values of the different parameters is given in Fig. \ref{Fig_DOS}. Note
the appearance of a peak and a discontinuity in the DOS, in the frequency range $\frac{\Delta}{2}<\hbar\omega<\frac{\Delta_\Gamma}{2}$, for $2mB\geq1$, signalling the change of regime pointed
above.
\section{Kubo formula: fixing  notation}
From linear response, we know that the average of the current operator is given by
\begin{equation}
 \langle J_\alpha\rangle = -\frac{i}{\hbar} \int_{-\infty}^tdt'
\langle [J_\alpha(t), V(t')]\rangle\,,
\end{equation}
where $V(t)$ is the perturbation. We consider an Hamiltonian, $H(\hbar\bm k)$,  such that
the electrons couple to the electromagnetic field through {\it minimal coupling},
that is, 
\begin{equation}
 H=H(\hbar \bm k-q\bm A)\,,
\end{equation}
where $q$ is the charge of the particles (for electrons we have $q=-e$, with $e>0$). 
The current is defined as
\begin{equation}
J_\gamma=-\frac{\partial H}{\partial A_\gamma}=\frac{q}{\hbar}
\frac{\partial H(\hbar\bm k)}{\partial k_\gamma}\,,
\label{eq_J_1st}
\end{equation}
and for a {\it small} $q\bm A$ (linear response) we have 
\begin{eqnarray}
 H(\hbar \bm k-q\bm A)&\approx& H(\hbar \bm k)-\sum_\gamma 
q A_\gamma\frac{\partial H(\hbar \bm k)}{\hbar\partial k_\gamma}
\nonumber\\
&=&H(\hbar \bm k) -\sum_\gamma J_\gamma A_\gamma,
\end{eqnarray}
thus, the perturbation reads
\begin{equation}
 V(t)=-\sum_\gamma J_\gamma A_\gamma(t)\,,
\end{equation}
where we have assumed that $A_\gamma(t)$ is a function of time. Then, the average of the 
$\alpha-$component of the current is 
\begin{equation}
 \langle J_\alpha\rangle = \sum_\gamma\frac{i}{\hbar} \int_{-\infty}^tdt'
\langle [J_\alpha(t), J_\gamma(t')]\rangle A_\gamma(t')\,,
\end{equation}
where $J_\alpha(t)$ is the current operator in the interaction picture.
We can now introduce a retarded function defined as
\begin{equation}
 \Pi^R_{\alpha\gamma}(t-t')=-i\theta(t-t')\langle [J_\alpha(t), J_\gamma(t')]\rangle\,,
\end{equation}
such that 
\begin{equation}
 \langle J_\alpha\rangle = \sum_\gamma\frac{-1}{\hbar}\int_{-\infty}^\infty
\Pi^R_{\alpha\gamma}(t-t')A_\gamma(t')\,.
\end{equation}
Fourier transforming the previous equation we obtain
\begin{equation}
 \langle J_\alpha(\omega)\rangle=-\frac{1}{\hbar}\sum_\gamma 
\Pi^R_{\alpha\gamma}(\omega)A_\gamma(\omega)\,.
\end{equation}
If we now choose
\begin{eqnarray}
 A_\gamma(t)&=&A_\gamma(\omega)e^{-i\omega t}
\nonumber\\
&\Rightarrow&
E_\gamma=-\frac{\partial A_\gamma(t)}{\partial t}=i\omega A_\gamma(\omega)e^{.i\omega t}\,,
\end{eqnarray}
thus $A_\gamma(\omega)=-iE(\omega)/\omega$. Finally, the conductivity tensor is 
\begin{equation}
\sigma_{\alpha\gamma}(\omega)=\frac{i}{\hbar\omega} \Pi^R_{\alpha\gamma}(\omega)\,.
\end{equation}
The retarded current-current correlation function is computed from the Matsubara
correlation function, which is defined as
\begin{equation}
 \Pi_{\alpha\gamma}(i\omega_n)=\int_0^{\beta\hbar} d\tau e^{i\omega_n\tau}\Pi_{\alpha\gamma}(\tau)\,,
\end{equation}
where
\begin{equation}
 \Pi_{\alpha\gamma}(\tau)=-\langle T_\tau J_\alpha(\tau)J_\gamma(0)\rangle\,,
\end{equation}
and $\beta=1/(k_BT)$, where $k_B$ is the Boltzmann constant and $T$ the temperature.
\section{Optical conductivity of a  two-band Hamiltonian: formal matters}

In connection with the Hamiltonian (\ref{eq_H0}),
let us now define creation and annihilation operators, $a^\dagger_{\bm k,\lambda}$
and $a_{\bm k,\lambda}$, which create and annihilate electrons in the band $\lambda$
with momentum $\hbar\bm k$; we denote $\lambda=+$ for the conduction band
and $\lambda=-$ for the valence band. In this basis, the current operators
in second quantization are defined as 
\begin{eqnarray}
 \hat J_x=\sum_{\bm k,\lambda,\lambda'}a^\dagger_{\bm k,\lambda}a_{\bm k,\lambda'}
\langle \psi_{\bm k,\lambda}\vert J_x\vert \psi_{\bm k,\lambda'}\rangle\,,\\
\hat J_y=\sum_{\bm k,\lambda,\lambda'}a^\dagger_{\bm k,\lambda}a_{\bm k,\lambda'}
\langle \psi_{\bm k,\lambda}\vert J_y\vert \psi_{\bm k,\lambda'}\rangle\,,
\end{eqnarray}
where $J_\alpha$, ($\alpha=x,y$) is defined in Eq. (\ref{eq_J_1st}).

In second quantization the  Matsubara current-current correlation function is written as
\begin{eqnarray}
\Pi_{\alpha\gamma}(i\omega_n)=-\int_0^{\beta\hbar}d\tau e^{i\omega_n\tau}
\sum_{\bm k_1,\lambda_1,\lambda'_1} 
\sum_{\bm k_2,\lambda_2,\lambda'_2}
\nonumber\\
\langle \psi_{\bm k_1,\lambda_1}\vert J_\alpha\vert \psi_{\bm k_1,\lambda'_1}\rangle
\langle \psi_{\bm k_2,\lambda_2}\vert J_\gamma\vert \psi_{\bm k_2,\lambda'_2}\rangle 
\nonumber\\
\times \langle a^\dagger_{\bm k_1,\lambda_1}a_{\bm k_1,\lambda_1}
a^\dagger_{\bm k_2,\lambda_2}a_{\bm k_2,\lambda'_2}
\rangle\,,
\end{eqnarray}
which in terms of Green's functions reads
\begin{eqnarray}
 \Pi_{\alpha\gamma}(i\omega_n)=-\int_0^{\beta\hbar}d\tau e^{i\omega_n\tau}
\sum_{\bm k_1,\lambda_1,\lambda'_1} 
\sum_{\bm k_2,\lambda_2,\lambda'_2}
\nonumber\\
\langle \psi_{\bm k_1,\lambda_1}\vert J_\alpha\vert \psi_{\bm k_1,\lambda'_1}\rangle
\langle \psi_{\bm k_2,\lambda_2}\vert J_\gamma\vert \psi_{\bm k_2,\lambda'_2}\rangle 
\nonumber\\
\times
{\cal G}(\bm k_1,\lambda_1,\tau)\delta_{\bm k_1,\bm k_2}\delta_{\lambda'_2,\lambda_1}
{\cal G}(\bm k_1,\lambda'_1,-\tau)\delta_{\bm k_1,\bm k_2}\delta_{\lambda_2,\lambda'_1}\,,
\end{eqnarray}
where
\begin{equation}
{\cal G}(\bm k,\lambda,\tau) =
-\langle T_\tau a_{\bm k,\lambda}a^\dagger_{\bm k,\lambda}(\tau)\rangle\,.
\end{equation}
Introducing the Fourier representation
\begin{equation}
 {\cal G}(\bm k,\lambda,\tau)  = \frac{1}{\hbar\beta}\sum_{i\omega_1}e^{-i\omega_1\tau}
{\cal G}(\bm k,\lambda,i\omega_1)\,, 
\end{equation}
we obtain
\begin{eqnarray}
\Pi_{\alpha\gamma}(i\omega_n)= \sum_{\bm k,\lambda,\lambda'} 
\langle \psi_{\bm k,\lambda}\vert J_\alpha\vert \psi_{\bm k,\lambda'}\rangle
\langle \psi_{\bm k,\lambda'}\vert J_\gamma\vert \psi_{\bm k,\lambda}\rangle 
\nonumber\\
\times
\frac{1}{\beta\hbar}\sum_{i\omega_2}
{\cal G}(\bm k,\lambda,i\omega_n+i\omega_2)
{\cal G}(\bm k,\lambda',i\omega_2)
\end{eqnarray}
which after summing over the Matsubara frequency $\omega_2$ gives
\begin{eqnarray}
 \Pi_{\alpha\gamma}(i\omega_n)= \sum_{\bm k,\lambda,\lambda'} 
\langle \psi_{\bm k,\lambda}\vert J_\alpha\vert \psi_{\bm k,\lambda'}\rangle
\langle \psi_{\bm k,\lambda'}\vert J_\gamma\vert \psi_{\bm k,\lambda}\rangle 
\nonumber\\
\times
\frac{n_F(\bm k,\lambda')-n_F(\bm k,\lambda)}{i\omega_n+(E_{\bm k,\lambda'}-E_{\bm k,\lambda'})/\hbar}
\,,
\end{eqnarray}
where $n_F(x)$ is the Fermi distribution function.
For $\lambda=\lambda'$ the previous result is zero due to the Fermi functions.
Finally, the retarded current-current correlation function is obtained from the 
Matsubara one by analytical continuation $\Pi^R_{\alpha\gamma}(\omega)
=\Pi_{\alpha\gamma}(i\omega_n\rightarrow \omega +i0^+)$.
Since 
\begin{equation}
\sigma_{\alpha\gamma}(\omega)=\frac{i}{\hbar\omega} \Pi^R_{\alpha\gamma}(\omega)\,,
\end{equation}
the real part of the conductivity tensor is given by 
\begin{equation}
\Re\sigma_{\alpha\gamma}(\omega)=\frac{i}{\hbar\omega}i\Im \Pi^R_{\alpha\gamma}(\omega)=
-\frac{\Im \Pi^R_{\alpha\gamma}(\omega)}{\hbar\omega}\,.
\end{equation}
From here on, we will be interested in the diagonal component of the conductivity
tensor.

\section{Optical conductivity of a modified Dirac equation}

As already noted, the retarded current-current correlation function $\Pi^R_{xx}(\omega)$ can be obtained
by analytical continuation of the Matsubara current-current correlation function, 
leading to an imaginary part of the form
\begin{eqnarray}
\Im\Pi^R_{xx}(\omega)= \frac{-\pi}{A}\sum_{\bm k,\lambda\neq\lambda'} 
\vert\langle \psi_{\bm k,\lambda}\vert J_x\vert \psi_{\bm k,\lambda'}\rangle\vert^2
\nonumber\\
\times
[n_F(\bm k,\lambda')-n_F(\bm k,\lambda)]
\delta[
\omega+(E_{\bm k,\lambda'}-E_{\bm k,\lambda'})/\hbar]
\,,
\label{eq_PI_xx}
\end{eqnarray}
where $A$ is the area of the system. The current operator $J_x$ is defined as
\begin{equation}
 J_x=-\frac{e}{\hbar}\sum_\alpha \partial_{k_x}d_\alpha(\bm k)\sigma_\alpha\,,
\end{equation}
which we use in the calculation of the matrix elements entering in Eq. (\ref{eq_PI_xx}). 
In the thermodynamic limit,
the momentum summation in Eq. (\ref{eq_PI_xx}) transforms into an integral 
in the usual way and one has to compute the angular average of the matrix elements, that is,
\begin{equation}
 I(k)=\int_0^{2\pi}d\theta \vert\langle \psi_{\bm k,\lambda}\vert J_x\vert \psi_{\bm k,\lambda'}\rangle\vert^2\,.
\end{equation}
The final result is $I(k)={\cal K}(k)\,,$
where
\begin{eqnarray}
 {\cal K}(k)&=&\frac{\pi\hbar^2}{E^2_{\bm k,+}}
[v^2(E^2_{\bm k,+}+M^2(\bm k))
\nonumber\\
&+&4BkM(\bm k)v\hbar\sqrt{E^2_{\bm k,+}-M^2(\bm k)}
\nonumber\\
&+&4B^2\hbar^2k^2(E^2_{\bm k,+}-M^2(\bm k))]\,.
\end{eqnarray}
The imaginary part part of the current-current correlation function reads
\begin{eqnarray}
\Im \Pi^R_{xx}(\omega)&=&-\frac{e^2}{\hbar}\pi\int_0^\infty
\frac{du}{8\pi^2}{\cal K}(u)\delta(\omega\hbar-2E_{u,+})\times\nonumber\\
&&[n_F(-\hbar\omega/2)-n_F(\hbar\omega/2)]\,,
\end{eqnarray}
where the change of variable $u=k^2$ was again made. In what follows, we assume that the
chemical potential lies in the energy gap and we take the zero temperature limit;
for finite temperatures, we have to keep the Fermi functions.
If we take the limit $m,B\rightarrow 0$, the real
part of the conductivity reads
\begin{equation}
 \Re\sigma_{xx}(\omega)=\frac{\sigma_0}{4}\,,
\label{eq_sigma_over_4}
\end{equation}
which is 1/4 the universal conductivity of neutral graphene, since we have not considered spin,
and in this case there is not a two-valley degeneracy as there is in graphene.
In the case $B=0$, the conductivity reads
\begin{equation}
 \Re\sigma_{xx}(\omega)=\frac{\sigma_0}{4}\left(
1+ \frac{m^2v^4}{\hbar^2\omega^2}
\right)\,,
\end{equation}
for values of $\hbar\omega$ greater than $2mv^2$. In this case, the optical
conductivity is at most $5\sigma_0/16$.

In the general case, of finite $B$ and $m$, we need again to evaluate 
the integration of the $\delta-$function to obtain $\Re\sigma_{xx}(\omega)$. The discussion is analogous to the
one made for the DOS, one merely needs to replace $\hbar\omega$ with $\hbar\omega/2$.
The analytical expression for the optical conductivity when both $B$
and $m$ are finite is too cumbersome to be given here and not much insight is gained.

\begin{figure}[ht]
\includegraphics*[width=8.5cm]{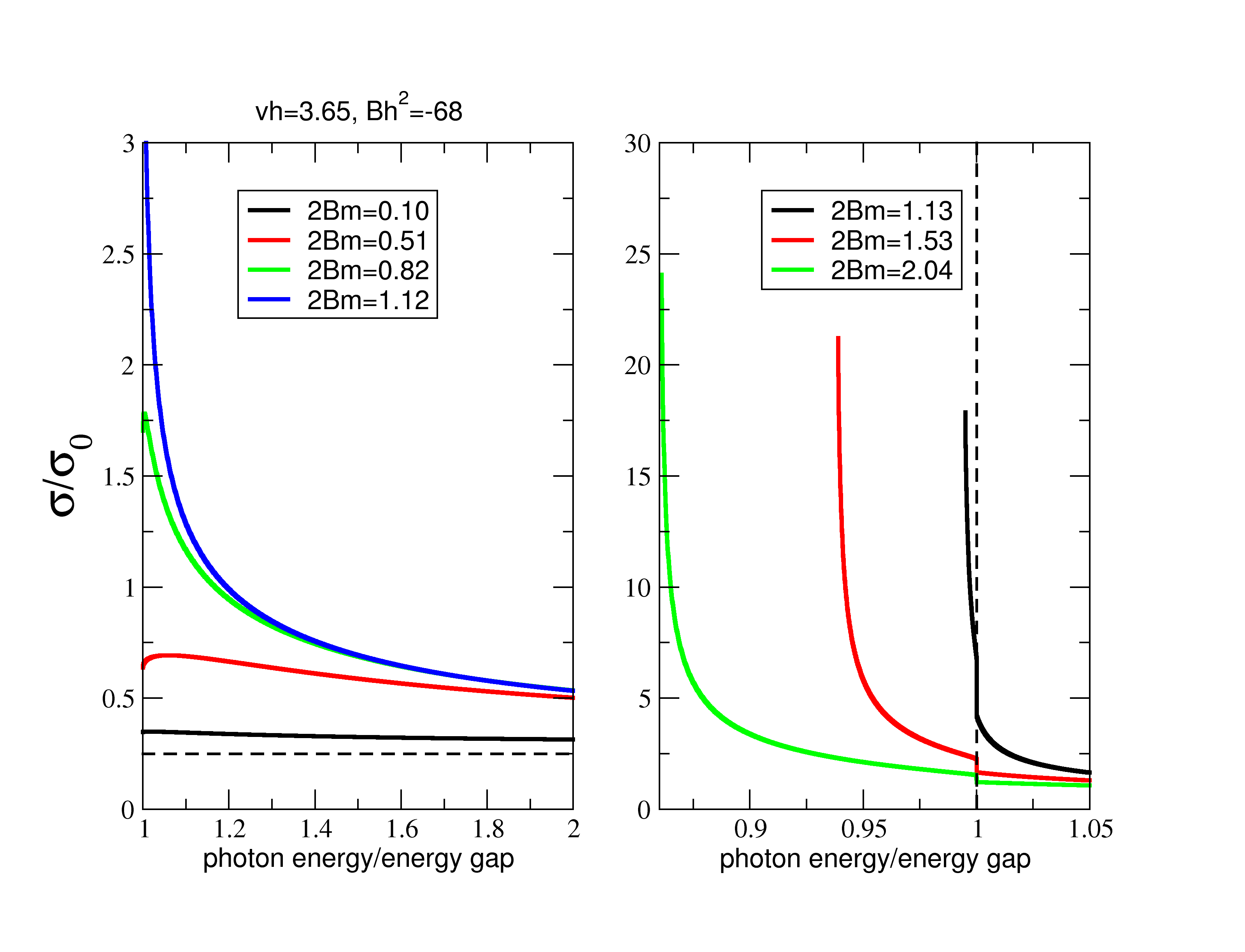}
\caption {Optical conductivity of the modified Dirac equation. The used parameters are
$v\hbar=3.65$ eV$\cdot$\AA, $B\hbar^2$=-68 eV$\cdot$\AA$^2$, as for 
HgTe quantum wells. The dashed line in the left panel is the result given by 
Eq. (\ref{eq_sigma_over_4}). The conductivity is in units of $\sigma_0$ and 
the photon energy in units of $\Delta_\Gamma=2mv^2$.}
\label{Fig_conductivity}
\end{figure}

In Fig. \ref{Fig_conductivity}, we plot the optical conductivity of the modified Dirac
equation. In the left panel, we follow the evolution of the optical conductivity 
upon the parameter $2Bm$. It is clear that as this parameter increases so does the 
optical conductivity, specially close to the gap edge. 
When $2Bm$ approach 1 the conductivity is greatly enhanced
close to the edge of the gap $\Delta_\Gamma$. When the system enters the regime $2Bm>1$ the 
conductivity can be enhanced by more than one order of magnitude for photon 
energies satisfying the relation
 $\hbar\omega\gtrsim\Delta$. Thus we find a strong light-matter interaction 
in the non-trivial regime for $2Bm>1$.
A particular feature of this regime
 is a  jump in the conductivity at $\hbar\omega=\Delta_\Gamma$. This jump correlates
with the same behaviour seen in the density of states and is a signature that the system
is in the non-trivial regime.

Finally, we have found that in the trivial regime the optical conductivity of the 
bulk system is of the order of magnitude as that measured for graphene.

\section{Conclusions}

We have discussed several aspects of the modified Dirac equation. We computed the
Chern number which defines the trivial and non-trivial regimes of the system.
We then computed the density of states and the optical conductivity of the modified
Dirac equation.
We found that in the non-trivial regime, characterized by $2Bm>1$, the 
density of states diverges as the  energy approaches  $\hbar\omega=\Delta$ and that the 
optical conductivity is greatly enhanced relatively to the case where $0<2Bm\ll 1$.
Indeed, the divergence in the density of states also appears in the optical conductivity
at photon energies close to $\Delta$. This divergence configures a strong light-matter
interaction for that range of frequencies. We then expect that physical effects such as
the Faraday rotation \cite{Faraday,EOM} must exhibit dramatic results, when compared to the case
of graphene, in the quantum regime dominated by inter-band transitions. 
Finally, we are confident that in the realm of  cold atoms the parameters $B$, $m$ and $v$
can be tuned at will, making possible the external tuning of the several regimes
and the observation of the different effects proposed here.

\ack
JES’s work contract is financed in the framework of the
Program of Recruitment of Post Doctoral Researchers for
the Portuguese Scientific and Technological System, with the
Operational Program Human Potential (POPH) of the QREN,
participated by the European Social Fund (ESF) and national
funds of the Portuguese Ministry of Education and Science
(MEC). The authors acknowledge support provided to the
current research project by FEDER through the COMPETE
Program and by FCT in the framework of the Strategic Project
PEST-C/FIS/UI607/2011.

\section*{References}

\begin{thebibliography}{99}

\bibitem{OpticalBook}
Fox M 2010 {\em Optical Properties of Solids\/} 2nd ed (Oxford)

\bibitem{rmp}
{Castro Neto} A~H, Guinea F, Peres N~M~R, Novoselov K~S and Geim A~K 2009 {\em
  Rev. Mod. Phys.\/} {\bf 81} 109

\bibitem{rmpPeres}
Peres N~M~R 2010 {\em Rev. Mod. Phys.\/} {\bf 82} 2673

\bibitem{SCZhang}
Qi X~L and Zhang S~C 2010 {\em Physics Today\/} {\bf 63} 33

\bibitem{Manoharan}
Manoharan H~C 2010 {\em Nature Nanotechnology\/} {\bf 477} 33

\bibitem{PWKane}
Kane C and Moore J 2011 {\em Physics World\/} {\bf 24} 32

\bibitem{TopBook}
Bernevig B~A 2013 {\em Topological Insulators and Topological
  Superconductors\/} (Princeton)

\bibitem{SCZhangNatPhys}
Zhang H, Liu C~X, Qi X~L, Dai X, Fang Z and Zhang S~C 2009 {\em Nature
  Physics\/} {\bf 5} 438

\bibitem{rmpKane}
Hasan M~Z and Kane C~L 2010 {\em Rev. Mod. Phys.\/} {\bf 82} 3045

\bibitem{rmpSCZ}
Qi X~L and Zhang S~C 2011 {\em Rev. Mod. Phys.\/} {\bf 83} 1057

\bibitem{carbote}
Li Z and Carbotte J~P 2013 {\em Phys. Rev. B\/} {\bf 87} 155416

\bibitem{nair}
Nair R~R, Blake P, Grigorenko A~N, Novoselov K~S, Booth T~J, Stauber T, Peres
  N~M~R and Geim A 2008 {\em Science\/} {\bf 320} 1308

\bibitem{StauberGeim}
Stauber T, Peres N~M~R and Geim A~K 2008 {\em Phys. Rev. B\/} {\bf 78} 085432

\bibitem{ZieglerTI}
Schmeltzer D and Ziegler K 2013 {\em arXiv:1302.4145\/}

\bibitem{BismuthBasov}
LaForge A~D, Frenzel A, Pursley B~C, Lin T, Liu X, Shi J and Basov D~N 2010
  {\em Phys. Rev. B\/} {\bf 81} 125120

\bibitem{Bismuth}
Pietro P~D, Vitucci F~M, Nicoletti D, Baldassarre L, Calvani P, Cava R, Hor
  Y~S, Schade U and Lupi S 2012 {\em Phys. Rev. B\/} {\bf 86} 045439

\bibitem{OpticalBi2Te2Se}
Akrap A, Tran M, Ubaldini A, Teyssier J, Giannini E, van~der Marel D, Lerch P
  and Homes C~C 2012 {\em Phys. Rev. B\/} {\bf 86} 235207

\bibitem{SPINShen}
Shen S~Q, Shan W~Y and Lu H~Z 2011 {\em SPIN\/} {\bf 1} 1

\bibitem{Faraday}
Ubrig N, Crassee I, Levallois J, Nedoliuk I~O, Fromm F, Kaiser M, Seyller T and
  Kuzmenko A~B 2011 {\em Nature Physics\/} {\bf 7} 48

\bibitem{EOM}
Ferreira A, Viana-Gomes J, Bludov Y~V, Pereira V, Peres N~M~R and {Castro Neto}
  A~H 2011 {\em Phys. Rev. B\/} {\bf 84} 235410

\end{thebibliography}



\end{document}